\documentclass[article]{aa}

\usepackage{graphicx}
\usepackage{txfonts}
\usepackage{hyperref}

\newif\ifdraft
\draftfalse   

\usepackage{dsfont}
\usepackage{amsmath}
\usepackage{amssymb}

\usepackage{xcolor}

\usepackage{ulem}      
\newcounter{tr}
\definecolor{darkgreen1}{rgb}{0.0,0.75,0.0}
\ifnum \value{tr}>5

\newcommand{\deleted}[1]{{\color{red} Deleted: } \sout{#1}}

\newcommand{\authorcomment}[1]{{\color{red} Comment :} {\color{Cyan} #1}}
\else

\newcommand{\deleted}[1]{}

\newcommand{\authorcomment}[1]{}
\fi

\begin{document}

\title{Automatic detection of long-duration transients in Fermi-GBM data}


\titlerunning{Long-duration transients}

\author{
	F. Kunzweiler \inst{1},
	B. Biltzinger\inst{1},
	J. Greiner \inst{1}
	\and J.M. Burgess \inst{1}
}
\institute{Max-Planck Institut f\"ur Extraterrestrische Physik, Giessenbachstrasse 1, 85740 Garching, Germany}

\date{Received 2022; accepted \_}

\abstract
{
In the era of time-domain, multi-messenger astronomy, the detection of transient events
on the high-energy electromagnetic sky has become more important than ever.
Previous attempts to systematically search for onboard-untriggered events in the data
of {\it Fermi}-GBM have been limited to short-duration signals with variability time scales
smaller than $\approx1$ min due to the dominance of background variations on longer timescales.
}
{
In this study, we aim at the detection of slowly rising or long-duration transient 
events with high sensitivity and full coverage of the GBM spectrum.
}
{
We make use of our earlier developed physical background model, which allows us to
effectively decouple the signal from long-duration transient sources from the complex
varying background seen with the {\it Fermi}-GBM instrument. 
We implement a novel trigger algorithm to detect signals
in the variations of the time series composed of simultaneous 
measures in the light curves of the different Fermi-GBM detectors in different energy bands.
To allow for a continuous search in the data stream of the satellite, the new
detection algorithm is embedded in a fully automatic data analysis pipeline.
After the detection of a new transient source, we also perform a joint fit for
spectrum and location, using the BALROG algorithm.
}
{
The results from extensive simulations demonstrate that the developed trigger 
algorithm is sensitive down to sub-Crab intensities (depending on the search 
time scale), and has a near-optimal detection performance.
During a two month test run on real {\it Fermi}-GBM data, the pipeline detected more than
300 untriggered transient signals.
For one of these transient detections we verify that it originated from a 
known astrophysical source, namely the Vela X-1 pulsar, showing pulsed emission for more than seven hours.
More generally, this method enables a systematic search for weak and/or long-duration transients.
}
{}


\maketitle
%
%
\section{Introduction}
Our understanding of the Universe was acquired throughout centuries of
observations of electromagnetic radiation and the study of cosmic rays.
While multi-messenger astronomy formally started with the detection of 
neutrinos from SN 1987A \citep{Arnett1989}, the era of multi-messenger 
time-domain astronomy was emerging in full glory with the detection of 
a binary neutron star coalescence GW170817 on 2017 August 17 and the 
gamma-ray burst GRB 170817A detected by {\it Fermi}-GBM and {\it INTEGRAL}-ACS 
\citep{Abbott_2017, Savchenko+2017}.
This new discovery confirmed the presumption that X-ray and $\gamma$-ray transients
play a crucial role in the development of multi-messenger astronomy.
In the past, it was demonstrated via sub-threshold analysis using BATSE 
\citep{Kommers+2001, Stern2002} and GBM \citep{Kocevski+2018} data 
that many GRBs stay undetected by the on-board trigger algorithms due to various effects,
including too week signal or too slow rising of flux.
This particularly applies to long-duration transients which rise more slowly than the trigger time scale used on-board.
Detecting such transient signals could significantly improve the understanding 
of source populations in the Universe, help multi-messenger astronomy (in case of faint, sub-threshold events), and 
enlarge the scientific outcome of satellite missions.

\begin{figure*}[ht]
    \centering
    \includegraphics[width=\linewidth]{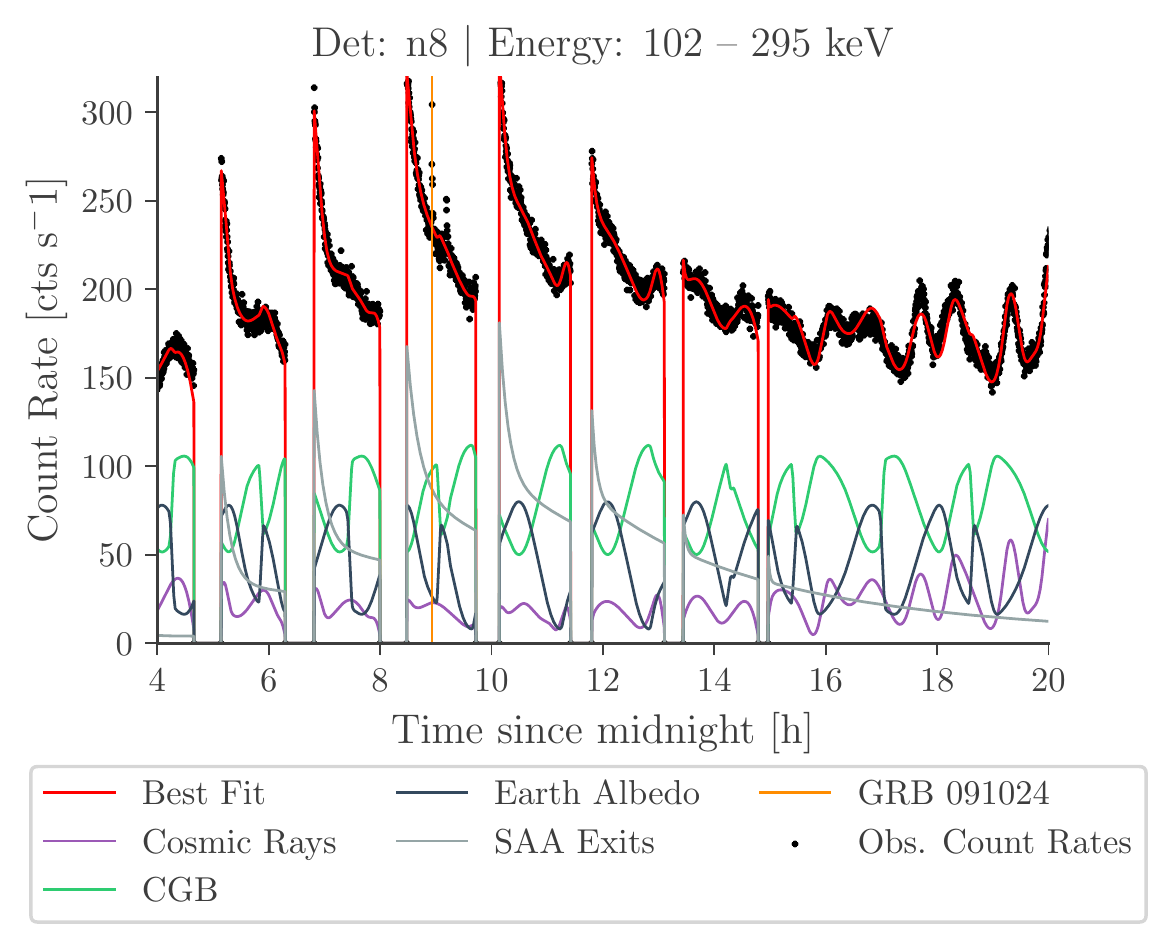}
    \caption{
    Data and background fit for the day of 24. Oct. 2009 for detector n8 in the reconstructed energy range of 
    102 -- 295\,keV with a time-bin size of 15\,s.
    This includes the ultra-long GRB 091024: its GBM trigger at 08:56:01 is marked by a cyan line
    \citep{Biltzinger2020}.
    }
    \label{fig:my_label}
\end{figure*}

All-sky monitoring instruments like {\it Fermi}-GBM are built to 
observe transient events in the $\gamma$-ray sky.
Detecting these transients is a non-trivial task as they can be concealed by a strongly varying 
background  (Fig. \ref{fig:my_label}) which becomes especially important for weak and long duration transients.
\textit{Swift}/BAT complements its trigger algorithm with a hard X-ray transient 
monitor\footnote{\href{ https://swift.gsfc.nasa.gov/results/transients/}{ https://swift.gsfc.nasa.gov/results/transients/}} \cite{Krimm_2013}.
In the case of {\it Fermi}-GBM, consisting of 12 NaI detectors and 2 BGO detectors and built 
primarily for detecting GRBs or similarly short-duration transients, two constraints come together:
(i) the background at low energies is composed by a variety of bright sources, so in order to 
avoid triggering on those, the onboard trigger algorithm omits the lowest two energy channels 
covering 4--12 keV and 12--27 keV \citep{Meegan+2009};
(ii) the temporal variations of the background in Fermi's low-Earth orbit have a time-scale of 
5--10 min, and consequently the longest onboard trigger timescale is set to 8.192 s \citep{Meegan+2009}: 
thus, slow-rising or long-duration transients cannot be detected.
The strongly varying background is the reason why there does not exist a pipeline searching for transients with 
variability slower than GRBs in the Fermi-GBM data. In all existing analysis tools, a polynom is fit to the background.
As soon as the rise time of a transient source gets slower than a few minutes, the polynomial fit cannot distinguish 
anymore between background and new transient source.
The recent work by \cite{Antier+2020} has tried to extend the covered energy range to 4 keV by
applying a median-smoothing filter to remove the variable background and thus allow the independent 
identification of gamma-ray bursts including the detection of soft X-ray long transients.
\cite{Antier+2020} also applied an upper limit of 50\,s to the allowed length of a transient signal, 
preventing its use for finding long duration transients.
This method has similar shortfalls as a polynomial fit as it is unable to distinguish between a new 
source, e.g. a long duration transient, and background variations.

With the advent of our earlier developed physical background model \citep{Biltzinger2020}
these limitations (both, at low-energy as well as long durations) can be circumvented,
motivating this work.
The detection of transient events and GRBs can be formulated as the statistical problem of detecting 
significant variations (hereby change-points) in the light curves (or more generally time-series).
The start- and endpoint of a transient are hereby estimated separately as two subsequent 
change-points in the time-series.
This method is traditionally not among the standard methods described in the GRB trigger literature 
\citep{Savchenko2012, Blackburn2015, Burns2019a}, but has recently been applied to the detection of GRBs in \cite{Antier+2020}.
Here, we describe our extensions of this concept of change-point detection by incorporating a physical 
background model and combining the information of multiple dimensions, i.e. the information of the 
different detectors and different energy bands per time step.

\section{Transient detection algorithm}

\subsection{Physical background model}
\label{section:phys_bkg_model}

For non-imaging instruments like GBM, the measured data are all-sky light curves with sources 
superimposed upon each other as a detector is scanning over the sky.
Thus, in any analysis of temporal transients such as GRBs, GBM relies on the ability to separate source
and background during any, or multiple, scan(s).
This is classically done by identifying off-source regions in time, i.e. time intervals of the data 
stream when the source is not active, or not in the field-of-view of a scanning detector),
and fitting smooth functions such as polynomials to each energy band's temporally evolving signal and 
extrapolating this model into the on-source region \citep{Pendleton1999,
  Greiner_2016}.
This common approach is only suitable for transient events, which are short with respect to the timescale 
of the typical temporal variation of the background (about 10 min in case of {\it Fermi}-GBM).
However, this leads to poor results for long duration events, as the extrapolation of the background can be 
inaccurate due to non-linear variations in the signal \citep{Biltzinger2020}.
This can be overcome by constructing a physical model for the photon background seen by {\it Fermi}-GBM and 
fitting the normalization parameters of the individual background components (Fig. \ref{fig:my_label}).
This predictive background model is designed with a minimum number of free parameters that incorporated
the physics of the source types, the response of the GBM detectors and the geometry and location of the spacecraft. 
This model should only fit the background and not also (long or slowly varying) transient signals, 
while allowing to extract the individual contributions of the different background sources.
The background components can be categorized in two main classes:
the photon components and the charged particle components.
The sources in the first class produce a photon spectrum, which is measured directly in the detectors.
The cosmic gamma-ray background, the Earth Albedo, point sources and the Sun are part of this class.
The second class is characterized by charged particles that alter the background via direct energy 
deposit in the detectors or as secondary radiation after interaction with the satellite material.
The origin of these charged particles are cosmic rays and those trapped 
in the South Atlantic Anomaly (SAA).
The physical background model has been derived and explained in detail
in \cite{Biltzinger2020}.

\paragraph{Parameter estimation}
These parameters of the physical background model can be inferred from the data using Bayesian methods.
\cite{Biltzinger2020} estimated these parameters using the nested sampling algorithm \texttt{MultiNest} 
\citep{Feroz2009} which provides good results for models with low to medium number of parameters.
The parameter estimation can made more robust by fitting the light curves of multiple detectors and energy bands jointly.
However, this increases the number of model parameters and the computational complexity significantly.
To overcome this, the background model has been implemented in the probabilistic programming language 
Stan which is state-of-the-art for statistical modeling and high performance statistical computation \citep{Carpenter2017}.
Full Bayesian inference is provided in Stan through two MCMC methods, the No-U-Turn sampler and an 
adaptive form of Hamiltonian Monte Carlo sampling \cite{Betancourt+2017.2}.
Given its advanced sampling methods, Stan is able to efficiently sample from very high dimensional 
posterior distributions, like our multi-detector and multi-energy band background fits.

\subsection{Change point detection}\label{section:cpd}
\label{methodology:pelt}

Many tasks in signal processing can be thought of as a search for an optimal partition of data 
measured over a time interval $I$.
One such task is the identification of a change in the underlying state of a data generating process, 
possibly several times in the given time interval.
These changes are commonly referred to as change-points and their detection has attracted 
researchers from the statistics and data mining communities for decades \citep{Basseville1995}.
Edge detection, segmentation, event detection and anomaly detection, are all similar methods and 
are occasionally applied to comparable problems
\citep{Aminikhanghahi2017}.
For our task of transient detection with GBM we use the data from the 12 NaI detectors
with 8 energy channels each leading to a total of 96 individual time series, 
i.e. a time series with 96 dimensions, motivating the use of a dimensionality reduction 
method shortly introduced in the following subsections.

\subsubsection{PELT - Pruned Exact Linear Time algorithm}
A naive approach to find the optimal partitions consists of finding the optimal segmentation by iterating 
over all possible segmentations of the signal and returning the one that minimizes the objective function.
The optimal partition algorithm introduced by \cite{Jackson2005} finds the global optimal segmentation of a time series 
with a computational cost of $\mathcal{O}(N^{2})$, the same as the greedy Bayesian Blocks algorithm.
\cite{Killick2012} recently extended the optimal partition
algorithm by introducing a pruning step within the dynamic program that
reduces the computational cost to linear speed $\mathcal{O}(N)$ with
mild assumptions, while not affecting the exactness of the resulting  solution,
hence the name \textbf{P}runed \textbf{E}xact \textbf{L}inear \textbf{T}ime (PELT).
PELT leads to substantially more accurate segmentation than Binary Segmentation 
\citep{Killick2012} while being asymptotically faster.

\subsubsection{High-dimensional change-point detection}
The previously discussed change-point detection algorithms work well in the uni-variate setting.
Their extensions to multiple dimensions quickly become computationally intractable with increasing dimensionality 
\citep{Scargle1998, Jackson2005, Fryzlewicz2014}.
This is commonly overcome by projecting the high-dimensional time-series to a lower dimensional 
one \citep{Horvath2012, Zhang2010,  Enikeeva2019}.
\cite{Grundy2020} introduced a new projection method that conserves changes in the first two moments and 
argues that the previous methods have been limited to detecting changes in a single parameter, usually the mean, 
which makes them impractical when multiple features of a time-series change.
This projection is performed by constructing a p-dimensional time series as a series
of p-dimensional column-vectors that are composed by the individual measurements 
at a time $t$.
Each vector is then mapped to two values, the distances and the separation angle 
to a reference vector.
For details on the choice of the reference vector and the mapping we refer to \cite{Grundy2020}.
As a result, by detecting changes in the distances 
and angles using a uni-variate change-point method like PELT,
the change-points in p-dimensional series can be recovered \cite{Grundy2020}.
\subsubsection{Application to {\it Fermi}-GBM}
\begin{figure}[ht]
    \centering
    \includegraphics[width=\linewidth]{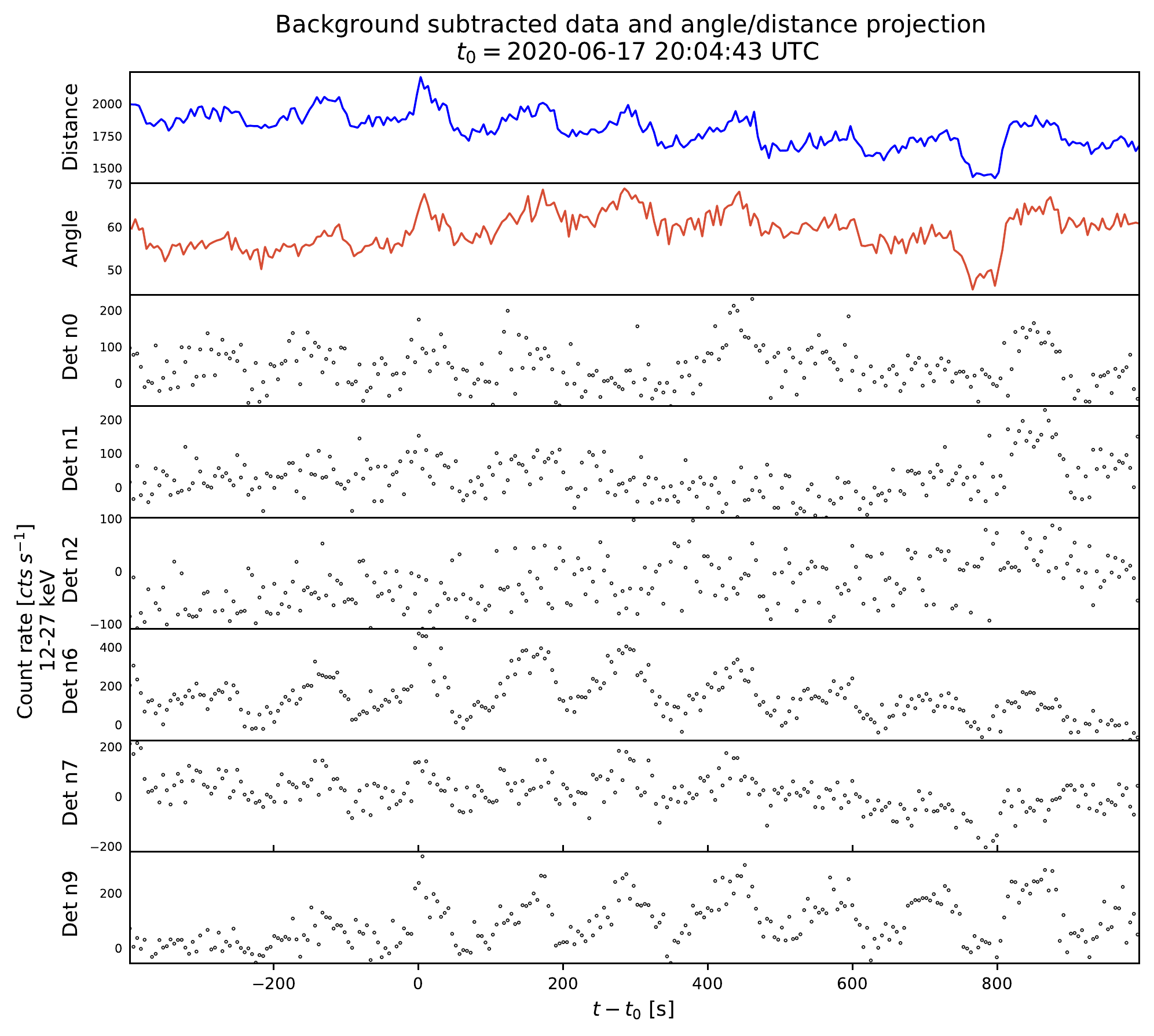}
	\caption{
    The background subtracted data of multiple detectors for the energy range 12--27 keV with a time binning of 5s
    in comparison to the projected angle and distance measure.
  }
    \label{fig:angle_distance_comparison}
\end{figure}

In the case of transient detection with GBM a transient signal comes from a certain direction and would 
therefore leave a signal in multiple detectors and energy channels depending on the spectrum and the location of the source.
This signal would be different in amplitude in the different detectors and would lead to a strong 
change in the angle measure and the distance measure shown in Fig. \ref{fig:angle_distance_comparison}.
In the case of a particle event there is a signal of similar strength in all detectors simultaneously,
which increases the distance measure but the angle would stay approximately constant.
This characteristic would make the angle measure more robust to false triggers on particle clouds.
Nonetheless, transient signals are a change in the mean of the time series and
investigation has shown that using both the distance and the angle measure
provides empirically the best trigger results.
Due to our aim of detecting soft and long-duration transients, we exclude the highest 
two energy bands from the trigger algorithm as they are dominated by background.

\subsection{Source detection}\label{section:source_detection}
Two consecutive change-points are combined as start and stop time to construct an
source interval.
For each source interval the significance is calculated by using the formula
introduced by \cite{Vianello2018} which applies some special treatment 
to avoid overestimating the significance of a candidate source by including the 
Gaussian error of the background fit performed using our model and applied to the unprojected data.
In this case the estimated background $b$ follows a Gaussian distribution $\mathcal{N}(B, \sigma)$, with
the true value $B$ and $\sigma$ as the error on the estimate.
Many practitioners including the recent transient search by
\cite{Antier+2020} determine the significance with the formula
$(n-b) / \sqrt{b}$, where $n$ is the number of observed counts.
This introduces the assumption that $n$ follows a Poisson distribution with
average $b$, which only holds if $B=b$; that the uncertainty on the background
estimate is negligible $\sigma \rightarrow 0$ and that $n$ is large enough to converge
to a Gaussian distribution with variance $\sqrt{b}$. 
Moreover, this form is a heuristically derived pseudo-statistic which has been shown to not map to 
an accurate significance distribution \cite{Li_Ma_1983}.
These assumptions are in most cases not justified and lead to an overestimated
significance.
The equation by \cite{Vianello2018} (i.e. Eq. 15) for the significance in case 
of a measurement with Poisson noise and a background with Gaussian noise is:
\begin{equation} \label{eq:significance}
  S = \sqrt{2} \Bigg[ n \log \bigg( \frac{n}{B_{0}^{mle}} \bigg) + \frac{(b-B_{0}^{mle})^{2}}{2 \sigma^{2}} + B_{0}^{mle} - n \Bigg]^{\frac{1}{2}}
\end{equation}
using the analytic solution $B_{0}^{mle}$ to the maximum likelihood estimate of the true background $B$.

For each source interval the significance is calculated for all detectors using
eq. \ref{eq:significance}.
Intervals with a significance of less than $5\,\sigma$ are discarded, whereas 
consecutive intervals with a significance higher than $5\,\sigma$ are combined.
Each source interval is then ranked by its significance in the most significant
detector.
When overlapping intervals are encountered the interval with a lower
significance is discarded.
\\

To determine the active time used in the following localization, we determine
for each detector the peak in excess counts and select an interval of $\pm 10$s
around that peak.
This time-interval has been selected as a trade-off between a longer
time-interval for increased statistics and the systematic localization error
introduced by the orbit of the satellite, i.e the boresight of Fermi 
(and thus that of most detectors) slewing with $1^{\circ}$ per 16s over the sky.
For these time intervals we calculate the significance and store the most
significant interval as active time for this trigger.

\subsection{Localization}\label{section:localization}
As {\it Fermi}-GBM is a non-imaging instrument, localizing transients is a challenging
task.
The applied method relies solely on the relative count rates observed by each
of its 14 detectors.
As the response of each GBM detector depends both on the incidence angle and 
energy of the detected photons, the measured count rate from a given transient 
source, depends on both the location and the physical spectrum of the source.
The ``BAyesian Location Reconstruction Of GRBs (BALROG)'' method, introduced by
\cite{Burgess+2018} and studied systematically by \cite{Berlato2019} fits 
simultaneously for both sky location and spectrum of a source.
This is done by constructing a full likelihood for the data set by combining the sky background from the physical background fit while propagating its errors,
the instrument detector response matrices (DRMs), and a model for the transient source.
The posterior samples are obtained by using \texttt{MultiNest} \citep{Feroz2009}, which implements a nested sampling algorithm.

\section{Automatic detection pipeline}

\begin{figure*}[!th]
    \centering
    \includegraphics[width=\linewidth]{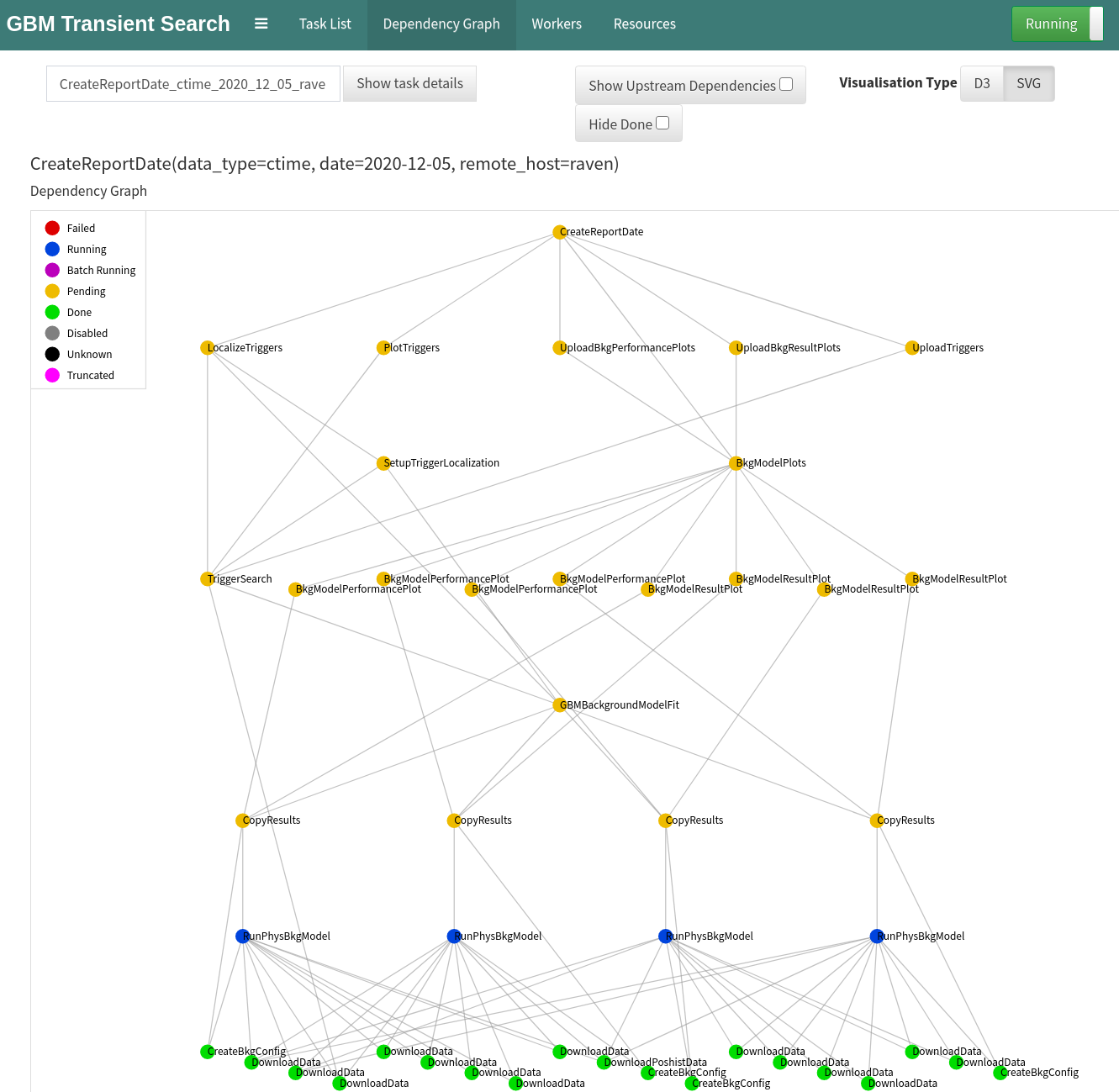}
    \caption{
    The dependency graph of the transient search pipeline, processing one day of data.
    Tasks that are already processed are marked with green circles, blue circles
    indicate tasks that are currently running and yellow circles indicate tasks
    that are pending due to missing dependencies.
    The processing starts at the bottom and the pipeline is currently running the fit of the
    background model ("RunPhysBkgModel"). 
    The following step "TriggerSearch" will run the change point detection (see subsection \ref{section:cpd}) and 
    source detection (see subsection \ref{section:source_detection}) of the transient detection algorithm.
    The subsequent "SetupTriggerLocalization" step will dynamically add individual tasks 
    for localization and spectral analysis (see subsection \ref{section:localization}) for every found trigger.
    All found transient signals are then uploaded to the GRB Science 
    Dashboard (\url{https://grb.mpe.mpg.de}).
    }
    \label{pipeline:grap}
\end{figure*}

To automatically search for transient signals, apart from the background
modeling, the transient detection algorithm and the source localization, one
key ingredient is still missing:
the individual processing steps and their dependencies have to be orchestrated
in a data analysis flow, that does not need human intervention.
This can be done by building a processing pipeline that combines multiple
processing steps and passes their outputs on to the next task.
The developed method is a fully autonomous data analysis pipeline built
with the luigi\footnote{\url{https://github.com/spotify/luigi}} framework,
that automatically downloads the satellite data as soon as they are available, 
runs all analysis steps and uploads a report to the GRB Science 
Dashboard\footnote{\url{https://grb.mpe.mpg.de}} (see Fig. \ref{pipeline:grap}).
The GRB Science Dashboard provides an interactive interface to the detected 
transient signals and their localization and spectral analysis.

\section{Simulations}

\subsection{Set-up}
In order to evaluate the performance of the background model fit and the transient search
algorithm, a simulation framework has been developed.
An artificial data file is created containing the expected background
fluctuations in the individual detectors and energy channels, by the following steps:
\begin{enumerate}
    \item Instantiate the background model components using the location history file of the satellite and the time bins of one day as described in \cite{Biltzinger2020}
    \item Keep the temporal evolution of the background components fixed
    \item Set the normalization parameters of the background model components to known (simulated) values
    \item Calculate the (time-dependent) flux of every component
    \item Fold each flux through the detector responses for each time bin
    \item Sum the contribution of all components
    \item Sample from the Poisson distribution with k equal to the sum of components for each time bin
    \item Store simulated count rates to FITS format like the original GBM data products
\end{enumerate}

By fitting the physical background model to this artificial data set, the
accuracy can be reviewed by comparing the posterior distribution over the
parameters with the real (simulated) values.
\\

To verify the performance of the trigger algorithm in a simulation, transient
signals can be superimposed on the background.
A realistic transient signal is simulated by defining a point source with a
physical spectrum and a position on the sky, and folding the point source flux
through a function defining the time evolution and thereby creating a transient
point source.
The flux of the transient point source is then folded through the detector
responses taking into account the Earth occultation to obtain the simulated
count rate per detector.
The simulated data-set is created by adding this transient count rate to a
simulated background model and subsequently adding Poisson noise.

In order to asses the sensitivity of the detection algorithm and the probability
to detect a signal of given duration and significance, two different transient
type light curves  have been studied by simulating $N=2x10^6$ transient signals,
as described in the next two subsections.

\subsection{A single-pulse type transient}

\begin{figure}[ht]
    \centering
    \includegraphics[width=\linewidth]{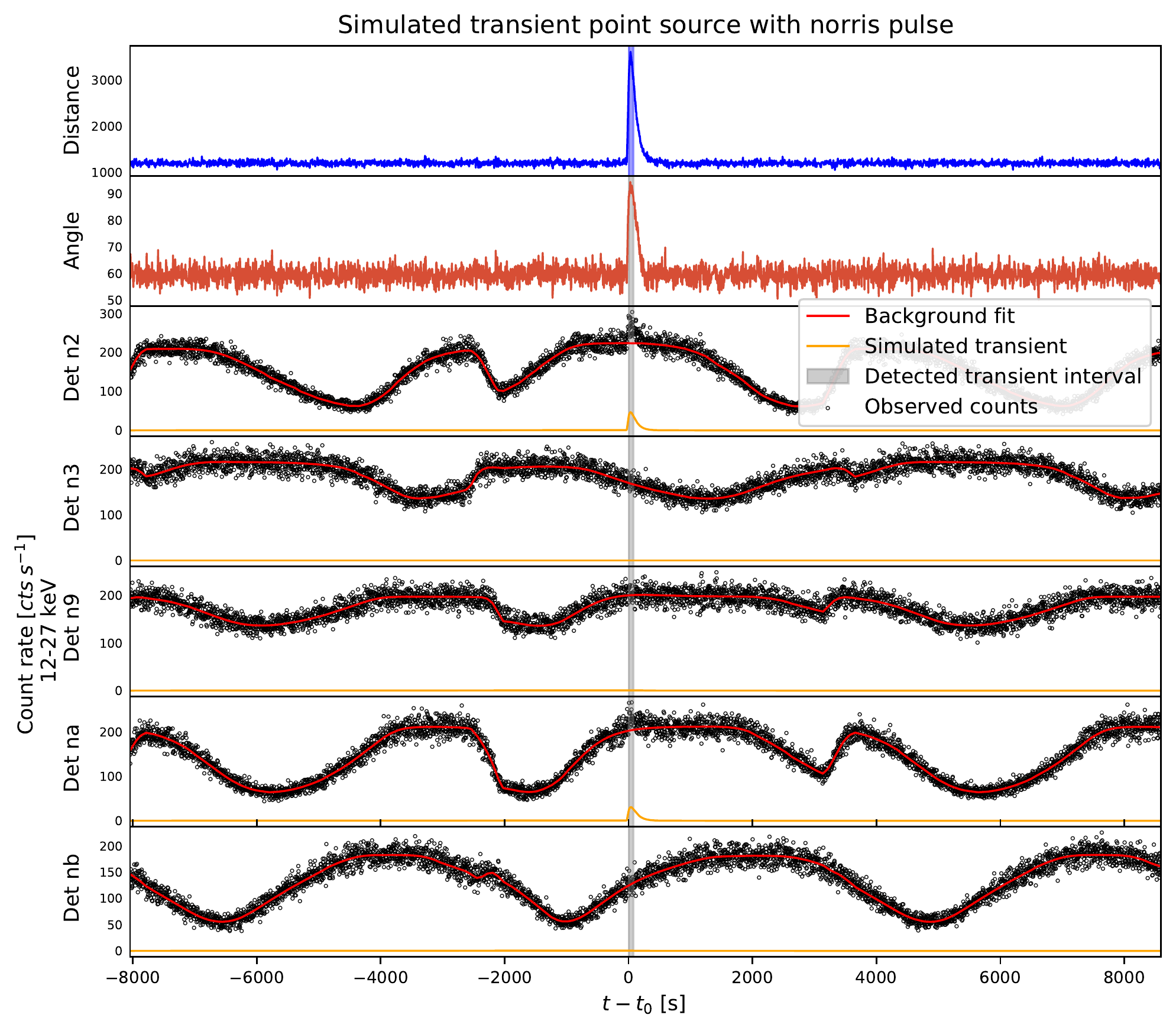}
    \caption{       
    Light curves of a simulated transient point source with at Norris pulse time 
    evolution on top of a simulated background.
    The count rate of the simulated transient is shown in orange, 
    the background fit in red and the time interval detected by the algorithm is highlighted in gray.
    The projection to the distance and angle measure is shown as a reference in the upper two subplots.
    }
    \label{fig:sim_transient_point_source}
\end{figure}

For single-pulse type transients a simulated source has been created with
a background-like spectrum (i.e. the sum of Earth Albedo and CGB given by
two broken-powerlaws 
with $\alpha^{Earth}_{1}=-5$, $\alpha^{Earth}_{2}=1.72$ and 
$\alpha^{CGB}_{1}=1.32$, $\alpha^{CGB}_{2}=2.88$ \citep{Ajello+2008}.
%
%
The start time $t_{start}$ and a location of the transient on the sky
have been sampled while making sure that the source was not occulted by the Earth.
As time evolution of the transient we have applied the Norris pulse function
\begin{equation}\label{eq:time_evolution_norris}
 f_{\textrm{norris}}(t) = A \cdot exp\Bigg(2\sqrt{\frac{t_{\textrm{rise}}}{t_{\textrm{decay}}}} \, \Bigg)  exp\Bigg(\dfrac{-t_{\textrm{rise}}}{t - t_{\textrm{start}}} - \frac{t - t_{\textrm{start}}}{t_{\textrm{decay}}}\Bigg),
\end{equation}
which has been demonstrated to be a good representation of GRB pulses
\citep{Norris1996}, by sampling values for $t_{rise}=t_{decay}$ from a Uniform distribution $\mathcal{U}\{0.1, 200\}$.
In Fig. \ref{fig:sim_transient_point_source} the simulated lightcurves of multiple detectors are shown for one simulation run.
The simulated transient events are then binned in two dimensions by 
their duration and significance.
This allows us to estimate the detection probability as
\begin{equation}
  P = \frac{n_{\textrm{detected}}}{N}
\end{equation}
for each bin individually and thereby as function of duration and significance
of the signal.
The error estimate for this probability can be obtained by using the frequentist formula
\begin{equation}
  err = Z  \sqrt{ \frac{ P (1 - P)} {N} }
\end{equation}
with $Z = 1.96$ for the 95\% error.

\begin{figure}[ht]
    \centering
    \includegraphics[width=\linewidth]{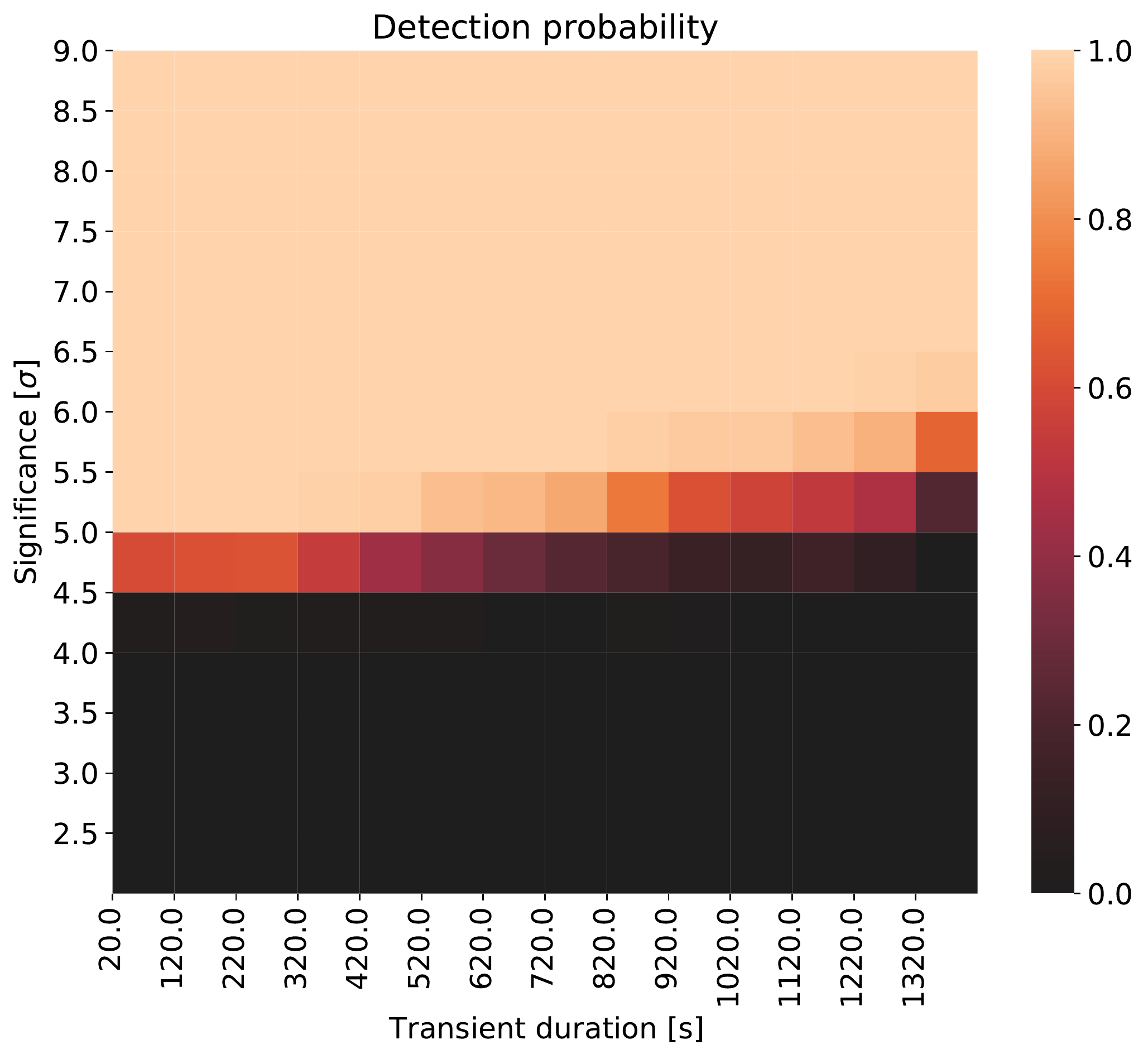}
    \caption{
    A 2D histogram showing the
    detection probability for a single-pulse type transient signal
    as function of its significance and duration.
    }
    \label{fig:sim_prob}
    \centering
    \includegraphics[width=\linewidth]{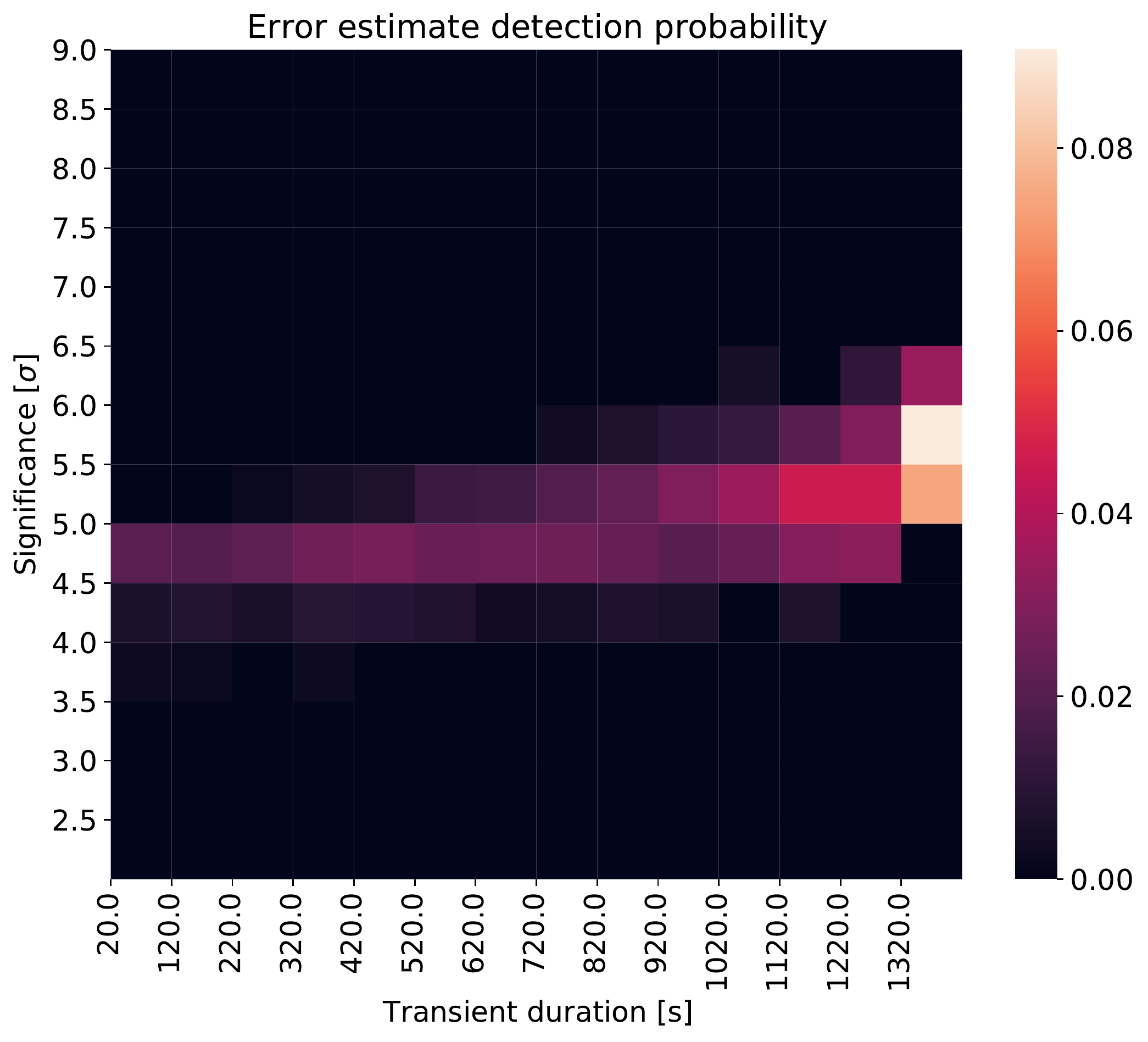}
    \caption{
    The frequentist error estimate of the detection probabilities shown in Fig. \ref{fig:sim_prob}
    }
    \label{fig:sim_prob_err}
\end{figure}

In Fig. \ref{fig:sim_prob} and Fig. \ref{fig:sim_prob_err} the detection probability 
and its error for signals with a duration of up to 1420\,s are visualized.
The view is clipped at a significance of $9\,\sigma$ for better visualization
as stronger signals were detected in 100\% of the trials.
For signals with a duration of 100\,s -- 200\,s and a significance between $4.5\,\sigma$
and $5\,\sigma$ we observe a detection probability of $P \approx 50\%$.
For the same durations we note for a significance of $< 4\,\sigma$ a probability of
$P \approx 0\%$ and for a significance of $>5\,\sigma$ a probability of
$P \approx 100\%$.
From this point the detection probability is reducing with
increasing signal duration for signals with a simulated significance close to
the $5\,\sigma$ threshold.

Although this finding is at first counter-intuitive, 
this can be explained by the calculation of the significance being influenced by the Poisson
noise especially in the tails of the pulse.
As the Norris Pulse has a smooth rise and decay we set the active time of the pulse to the
time interval where the pulse has an amplitude $f(t) > 0.1\,A$, calculated the significance 
for this interval and assumed this as the simulated significance.
Yet, it can lead to a higher significance than the simulated value, when the algorithm
selects less of the peak, which increases the detection probability below the
$5\,\sigma$ threshold.
At the same time long duration signals with just above $5\,\sigma$ of
simulated significance are smooth and consist of a slow rise and decay.
The algorithm will detect the start of the transient with a delay for smooth signals,
whereas the end of the signal is detected too early.
Therefore, the method will not integrate over the entire signal, which can lead
to a sub-threshold significance, and therefore to a neglected detection.
This is balanced out by long duration signals with higher significances as they
still pass the $5\,\sigma$ threshold even if we integrate only over some portion
of the signal.
\\

\begin{figure}[ht]
    \centering
    \includegraphics[width=\linewidth]{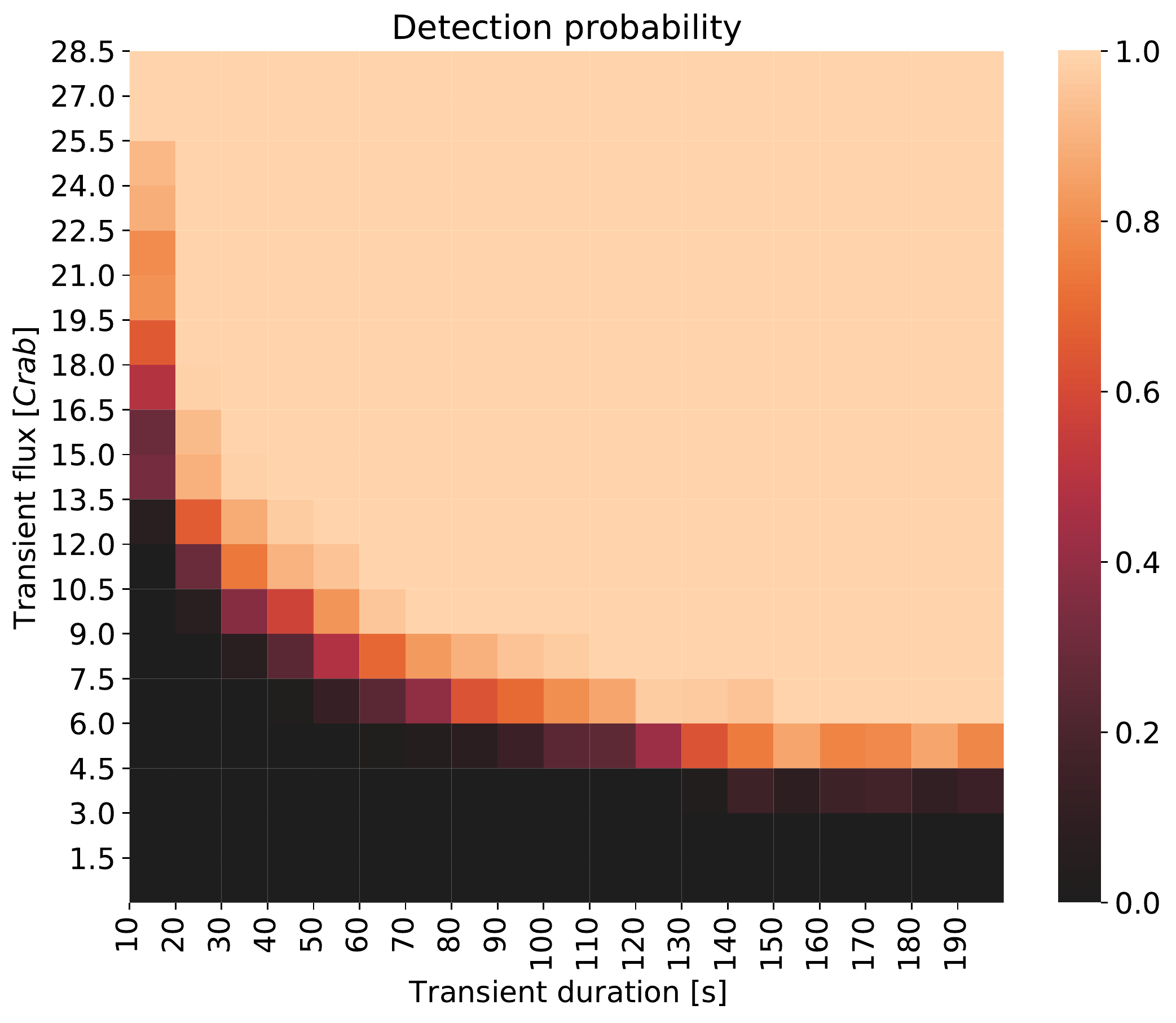}
    \caption{    
    A 2D histogram showing the detection probability for a single-pulse type transient signal
    as function of the transient flux in units of mCrab and its duration for short duration
    signals between 10s and 200s.}
    \label{fig:sim_prob_flux_lim_short}
    \centering
    \includegraphics[width=\linewidth]{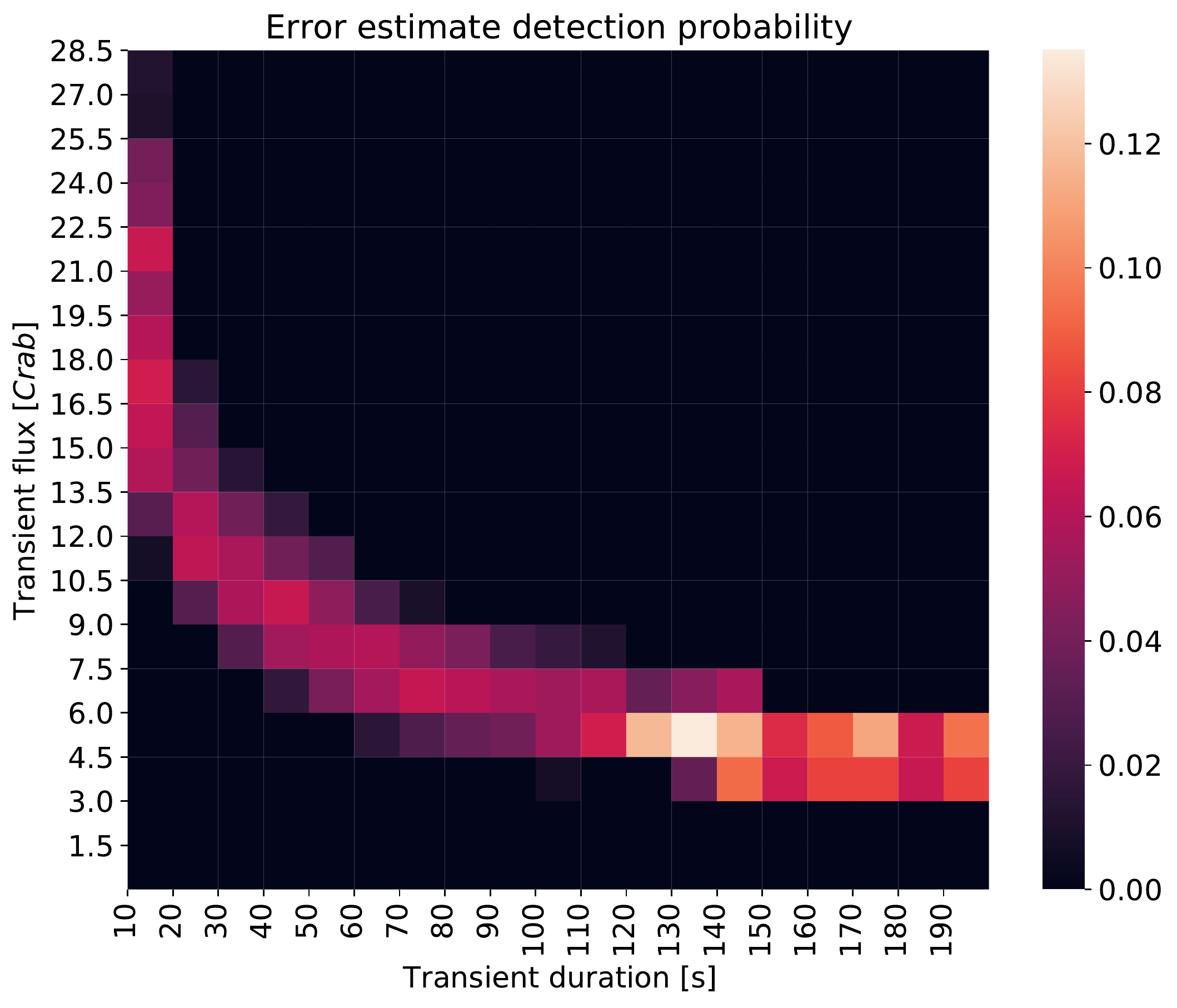}
    \caption{       
    The frequentist error estimate of the detection probabilities shown in
    Fig. \ref{fig:sim_prob_flux_lim_short}
    }
    \label{fig:sim_prob_flux_lim_err_short}
\end{figure}

\begin{figure}[ht]
    \centering
    \includegraphics[width=\linewidth]{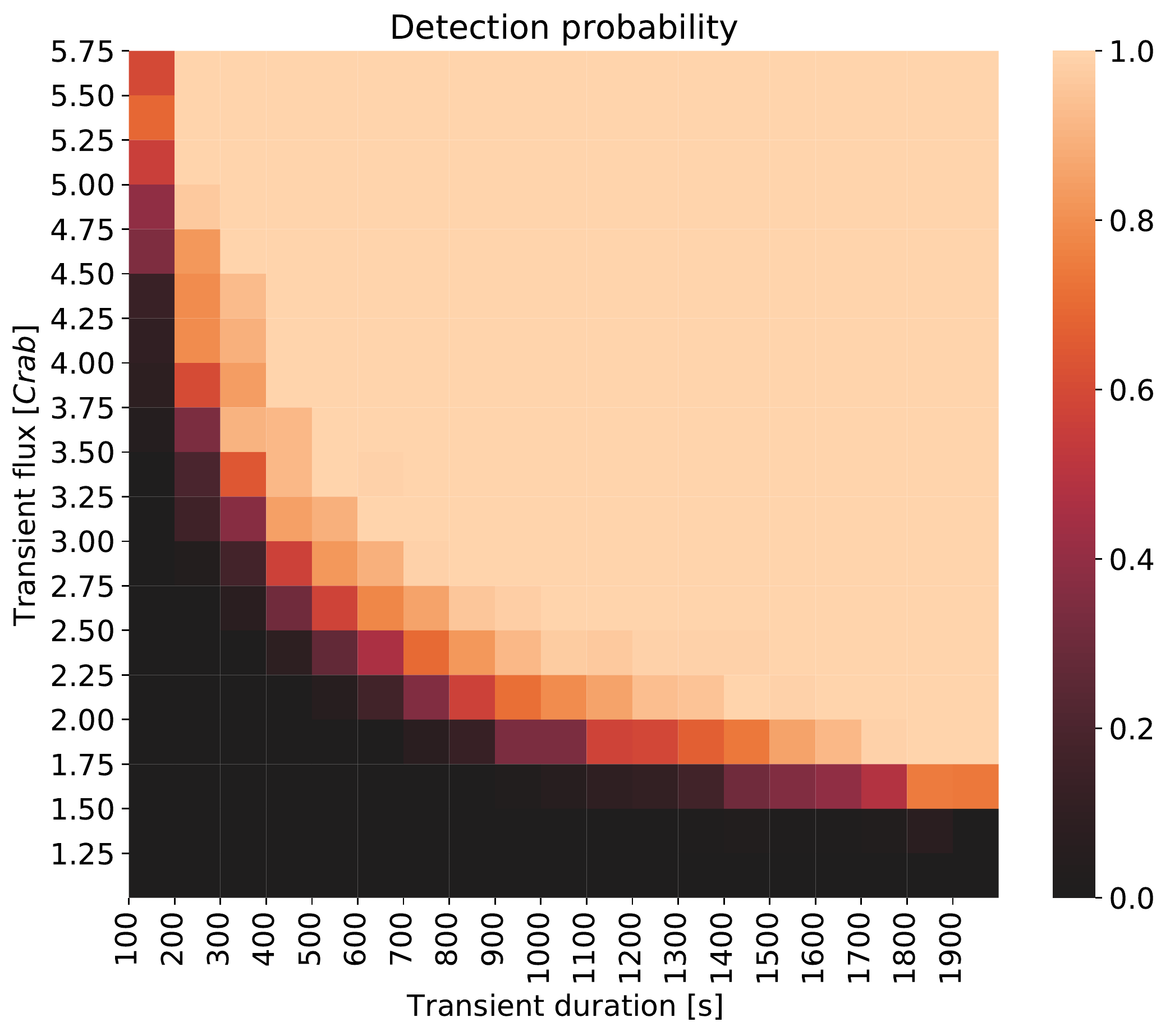}
    \caption{A 2D histogram showing the detection probability for a single-pulse type transient signal as
    function of the transient flux in units of mCrab and its duration for long duration 
    signals between 100\,s and 2000\,s.}
    \label{fig:sim_prob_flux_lim_long}
    \centering
    \includegraphics[width=\linewidth]{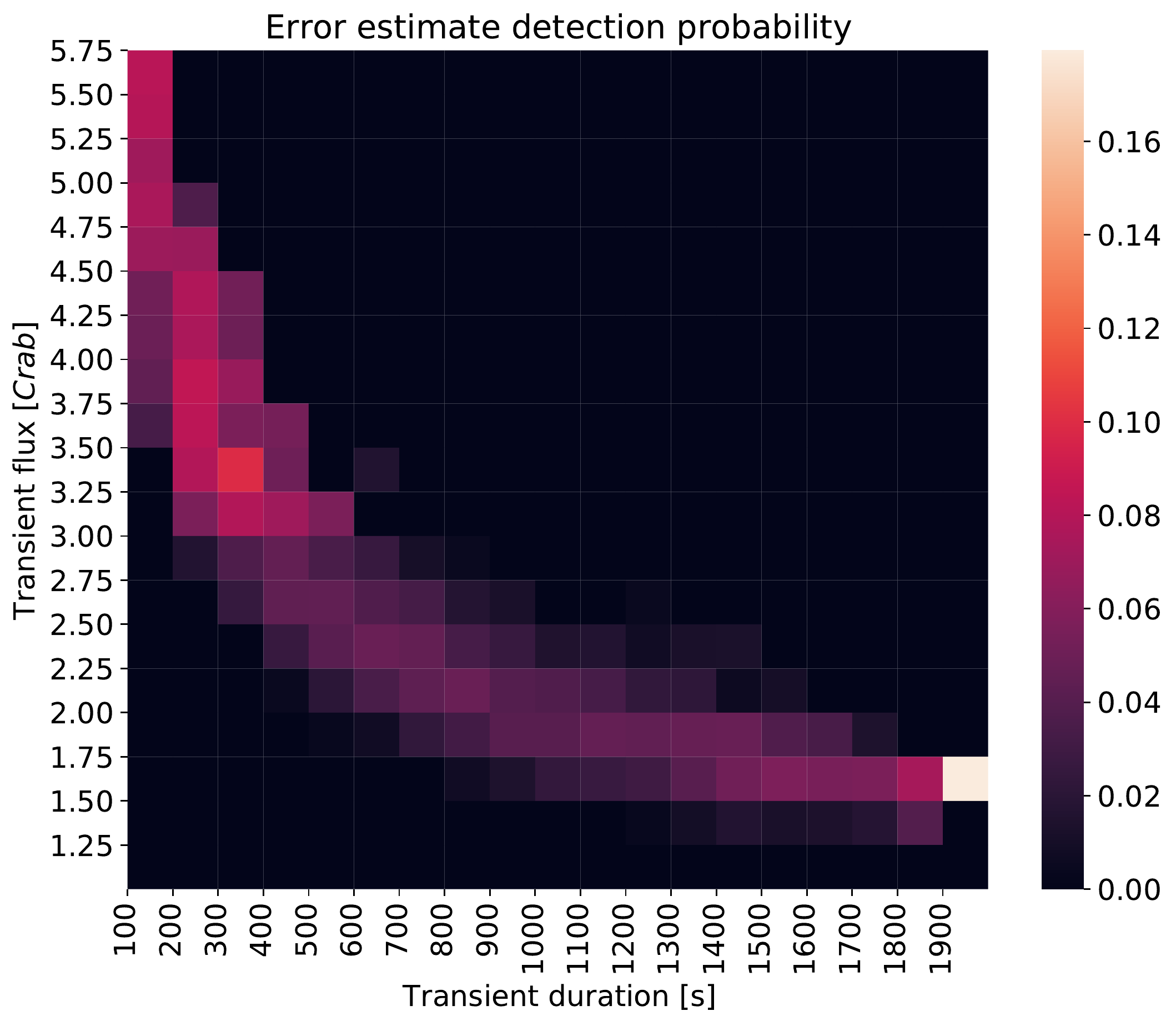}
    \caption{The frequentist error estimate of the detection probabilities shown in
    Fig. \ref{fig:sim_prob_flux_lim_long}
    }
    \label{fig:sim_prob_flux_lim_err_long}
\end{figure}

To estimate the detection probability at a given flux of the transient, we will
simulate the transient with the spectrum of the Crab nebula, a power-law
%
with a spectral index of $2.106$ and a normalization of
$9.71 \; {\rm keV^{-1}cm^{-2}s^{-1}}$
\citep{Madsen2017}.
The estimated detection probability plotted as a function of source flux in unites of Crab
and the duration of the simulated transient is illustrated in 
Fig. \ref{fig:sim_prob_flux_lim_short} and Fig. \ref{fig:sim_prob_flux_lim_err_long}.
It is evident that with increasing transient duration, the required flux for a 
detection is decreasing.
A detailed view for short duration signals is displayed in Fig.
\ref{fig:sim_prob_flux_lim_short} and Fig. \ref{fig:sim_prob_flux_lim_err_short}.
It shows that for short duration signals of 10--25\,s a $5\,\sigma$ detection requires
a flux of up to 25 Crab.
This could be at first surprising and should not be confused with the GRB detection
sensitivity, but the reason for this is twofold:
firstly, the time binning of the light curves in this simulation
was 5\,s, which as previously shown leads to a decrease in detection 
for signals with a duration of only a few time bins due to non-optimal time-selection.
secondly, the significance of the detected signals is calculated taking into
account the uncertainties of the physical background model, which slightly
reduces the significance of a detection.

To put this flux into context we will compare it to the short GRB 170817A, which
is considered to lay close to the detection threshold of GBM.
The reported peak flux in the energy range of 10-1000 keV of this GRB was
$(7.3 \pm 2.5) \times 10^{-7} \; {\rm erg \ s^{-1} cm^{-2}}$ \citep{Goldstein2017}.
This can be converted in units of Crab, which corresponds to a flux of
$(14.4 \pm 4.9) \textrm{ Crab}$.
A short GRB of 10\,s duration with the same flux as GRB 170817A would be detected
with a 35\%  probability with our method optimized for long duration signals.
It is likely that this can be further improved by using a finer time binning,
which we leave for future optimization.

The detection threshold for long duration signals is illustrated in Figure
\ref{fig:sim_prob_flux_lim_long}.
With increasing transient duration the flux required for a detection is decreasing.
This is expected, because integration over long duration signals with the same flux
leads to a higher significance than for short duration signals.

\subsection{A switching-on transient}

One potential application of the method developed in this work, is the detection
of an astrophysical source going into outburst over a long period of time, which
can last over multiple days.
For simplification we assume the source to have a constant flux during the
active time.
Typically the flux of these sources is reported in units of the Crab nebula
which is considered a standard candle in high-energy astrophysics.
Therefore, we set the spectrum of the point source in this simulation to the
spectrum of the Crab.

\begin{figure}[ht]
    \centering
    \includegraphics[width=\linewidth]{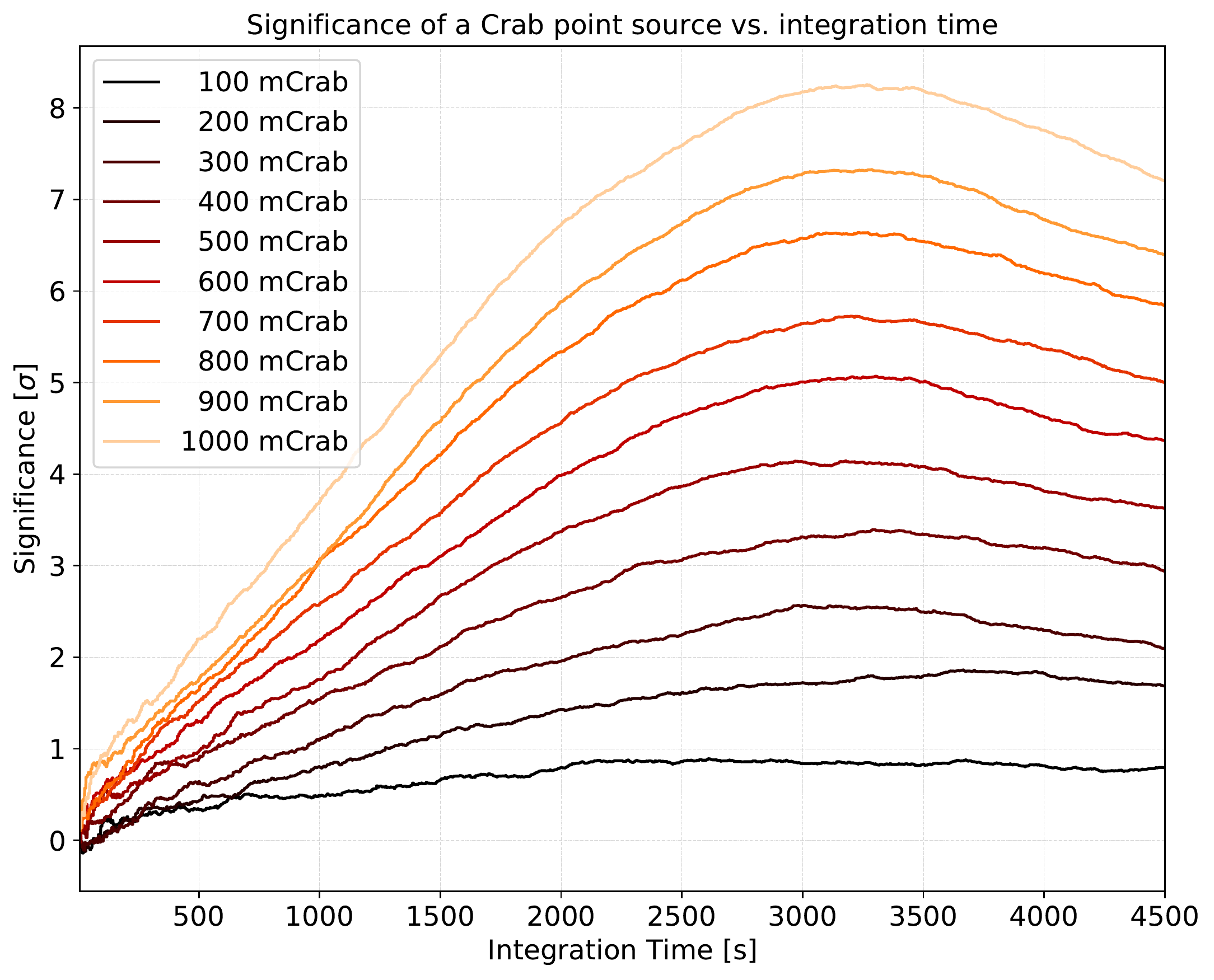}
    \caption{       
    An illustration of the significance of a switching-on transient point source with a Crab
    spectrum as a function of integration time.
    The used detector sweeps over the point source with its optical axis at
    1520\,s.
    The start time of the integration has been chosen to maximise the overall
    significance.
    This represents the theoretical limit of significance that can be measured
    for this type of source, correctly incorporating the error of the background estimate.
    }
    \label{fig:sim_ps_int_time}
\end{figure}

To understand the required flux from this point source it is useful to first
look at the ideal case, where one detector sweeps directly over the point source
with its optical axis.
In Fig. \ref{fig:sim_ps_int_time} the measured significance in this detector is
shown as a function of the integration time.
The start time of the integration has been set, by maximising the significance
as a function of start time and integration time, and choosing the optimal
value.
The detector passes over the point sources with its optical axis at $t=1520$\,s and
the maximum significance is reached at an integration time of approximately $3200$\,s.
After this time the significance decreases as the point source is not visible by
this detector anymore and we start integrating over background only.
The peak differs slightly for each simulated curve due to the Poisson noise
introduced in the simulated count rate.
This shows that for a $5 \sigma$ detection we need a point source with at least 600
mCrab and integrate over more than 3000\,s.
A point source with 500 mCrab does not reach the $5 \sigma$ threshold even if we
integrate over the entire visible orbit of approximately 3200\,s.

\begin{figure}[ht]
    \centering
    \includegraphics[width=\linewidth]{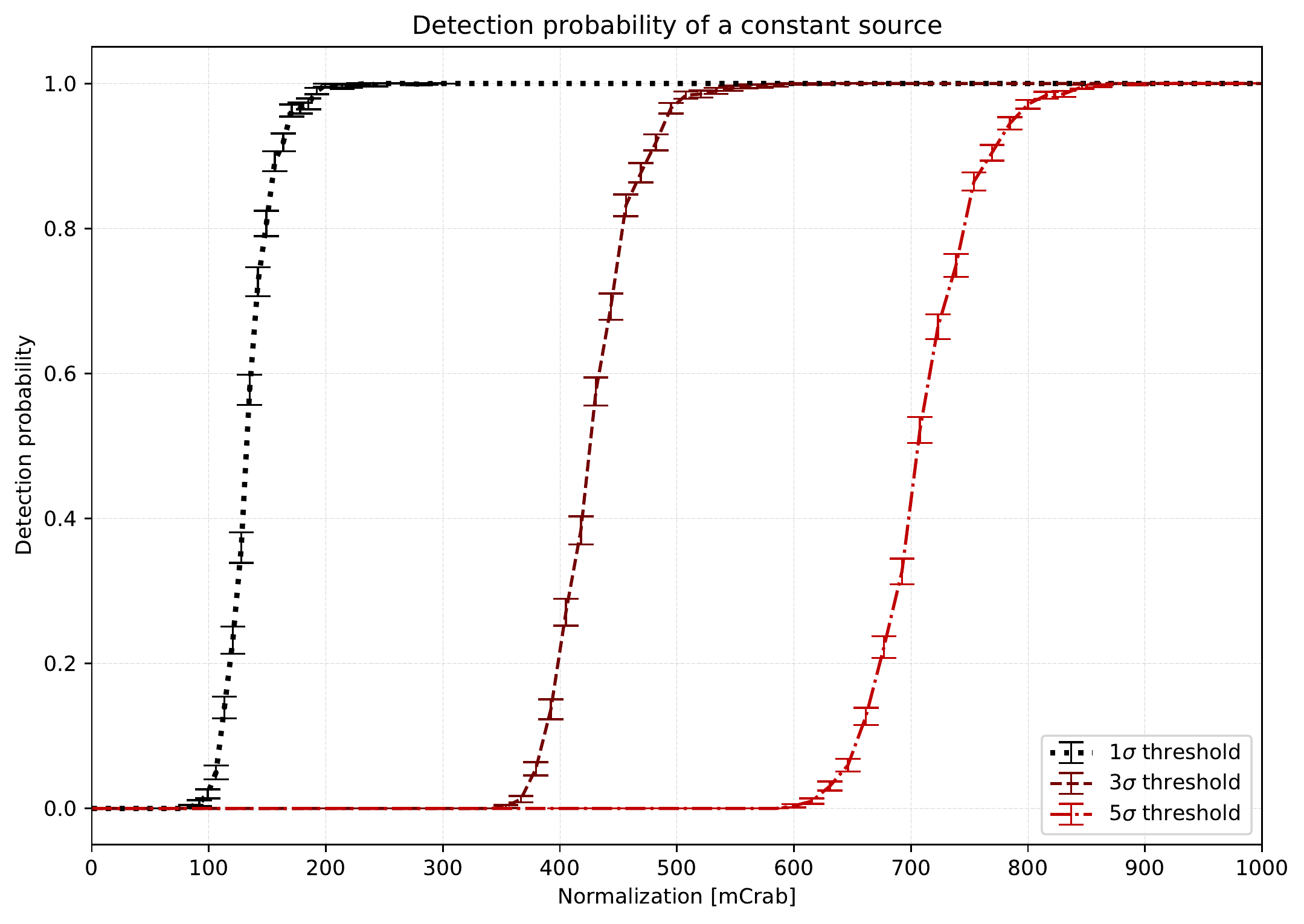}
    \caption{       
    Detection probability for a switching-on transient source with a Crab
    spectrum as a function of flux at a detection threshold of $1\sigma$, $3 \sigma$ and $5 \sigma$.
    }
    \label{fig:sim_prob_flux_lim_err_crab}
\end{figure}

To estimate the ability to detect this type of transient event with the
automatic detection algorithm, an additional simulation has been performed.
A transient point source that ``turns on'' instantly, stays constant over 24
hours and then stops stops instantly is simulated.
Throughout the simulation the normalization has been sampled uniformly between 
10 mCrab and 10 Crab.
To marginalize over the detector configuration the position of the point sources
has been sampled from a disc with 50 degree diameter.
The detection probability can be calculated as introduced in the previous
section by binning the simulated transients by the normalization and estimating
the detection probability as $P = n_{\textrm{detected}} / N$.

The detection probability as a function of the point source flux is shown in
Fig. \ref{fig:sim_prob_flux_lim_err_crab} for a required detection significance of
one sigma, three sigma, and five sigma.
The algorithm shows a close to optimal performance, as the detection probability
increases strongly starting at a flux level of 600 mCrab and reaches more than
95\% at 800 mCrab.
It has to be highlighted that during this simulation the location of the
transient source is sampled randomly.
Therefore, for only a fraction of simulated point sources a detectors sweeps
over the source location on-axis.
Furthermore the transient detection algorithm has not been tuned for this
specific simulation but detected the start and integration time via the
introduced change point detection method, running the (automated) detection pipeline.
As we lower the required significance threshold we are able to detect events with a lower flux.

\section{Physical Transients}

To verify the performance of the pipeline with real data, two test runs have been
performed spanning over approximately two months (2020/06/16 -- 2020/07/14 and
2020/10/24 -- 2020/11/17). 
The pipeline was run on chunks of data-sets of 1 full day, and the processing time per day was about 5 hrs.
During this time the pipeline has run autonomous without a human in the loop.
In this sample run the algorithm has successfully detected and localized more than 300 transient signals. 
A quick inspection revealed that only a handful of triggers could be classified as questionable.
Using triggers with a $2\,\sigma$ localization error smaller than 25 deg (about 50\% of all triggers),
we find 12\% of the triggers to be consistent with solar flares, 
60\% coming from known point sources such as the Crab, bright X-ray binaries such as Sco X-1 
(dominatinating) or Cyg X-1, and 28\% having no obvious high-energy counterpart in their error circle.
Identifying these latter sources in the future will be a worthwhile subject of research.

\begin{figure}[ht]
    \centering
    \includegraphics[width=\linewidth]{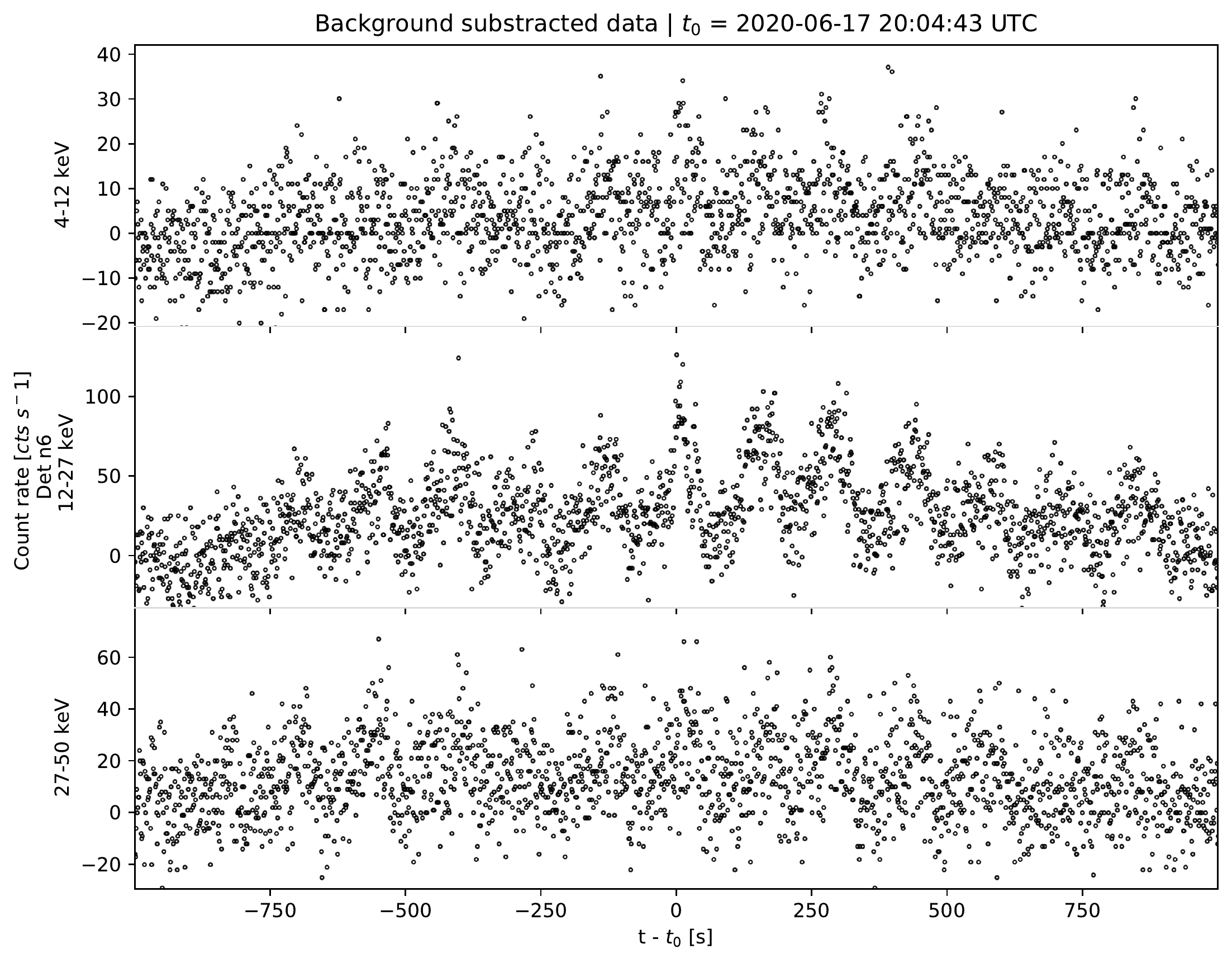}
 	\caption{Observed data of GBM detector n6 around the time of the trigger.}
     \label{fig:velax1_trigger_n6}
\end{figure}

On June 17, 2020, the transient search pipeline triggered three times on a
pulsating signal.
\begin{enumerate}
  \item 19:48:35 UTC with a duration of 665.6\,s and a significance of $12.9\,\sigma$.
  \item 20:04:53 UTC with a duration of 1593.6\,s and a significance of $33.4\,\sigma$ (see Fig. \ref{fig:velax1_trigger_n6}).
  \item 20:21:42 UTC with a duration of 153.6\,s and a significance of $5.3\,\sigma$.
\end{enumerate}
The automatic localization of the these three signals coincided at the location
of $\textrm{RA} = 142.2 {\rm \;deg}$ and $\textrm{DEC}=-37.7 {\rm \;deg}$ with a 
$2\,\sigma$ error circle of 56.4 degrees.
To demonstrate the the capacity of this transient search method to find real
physical signals, a manual analysis of this event has been performed to find a
possible counterpart.
In Fig. \ref{fig:velax1_trigger_n6} the background subtracted data of detector
n6 is shown around the trigger time of the second (strongest) trigger for the lowest
three energy channels.
A clear oscillating signal is visible with a periodicity of $T \approx 142\,s$.

\begin{figure}[ht]
    \centering
    \includegraphics[width=\linewidth]{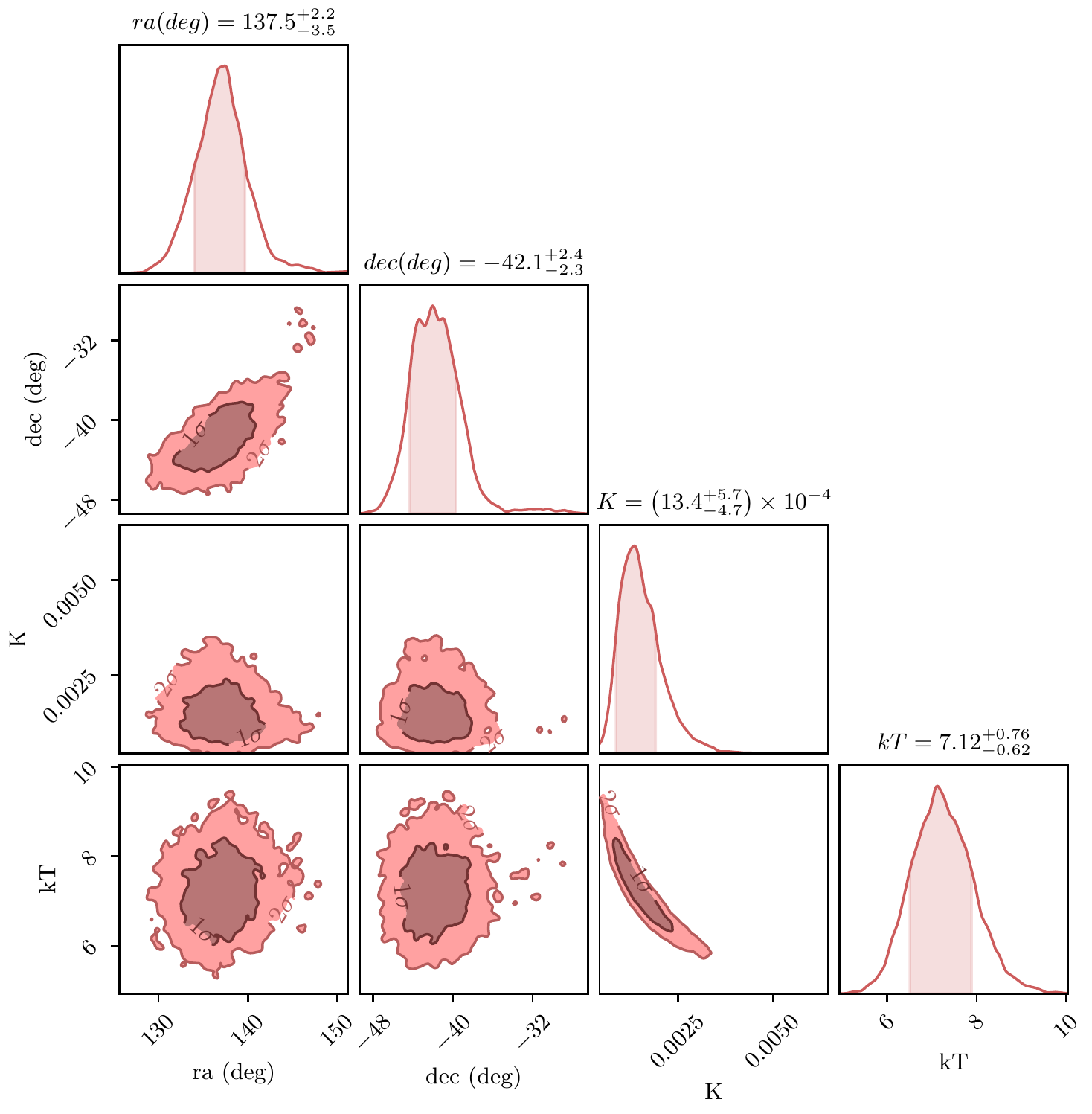}
	\caption{
    Result of the BALROG fit of the oscillating source with a blackbody spectrum
    using the data from multiple peaks.
  }
	\label{fig:velax1_balrog_corner}
\end{figure}

In order to improve the error of the automatic localization of the pipeline, a
manual localization has been performed.
Multiple active time intervals have been selected covering multiple peaks, and 
for each interval a response is generated using \texttt{gbm\_drm\_gen} \citep{Burgess+2018}.
For each interval the count spectrum is generated and a joint BALROG fit is
performed assuming a blackbody spectrum for the source, while linking the
position parameters.
In Fig. \ref{fig:velax1_balrog_corner} the pair plot from this fit is shown,
constraining the location of this source to $\textrm{RA}=137.5^{+2.2}_{-3.5} {\rm \;deg}$ and
$\textrm{DEC}=-42.1^{+2.4}_{-2.3} {\rm \;deg}$.

\begin{figure}[ht]
    \centering
    \includegraphics[width=\linewidth]{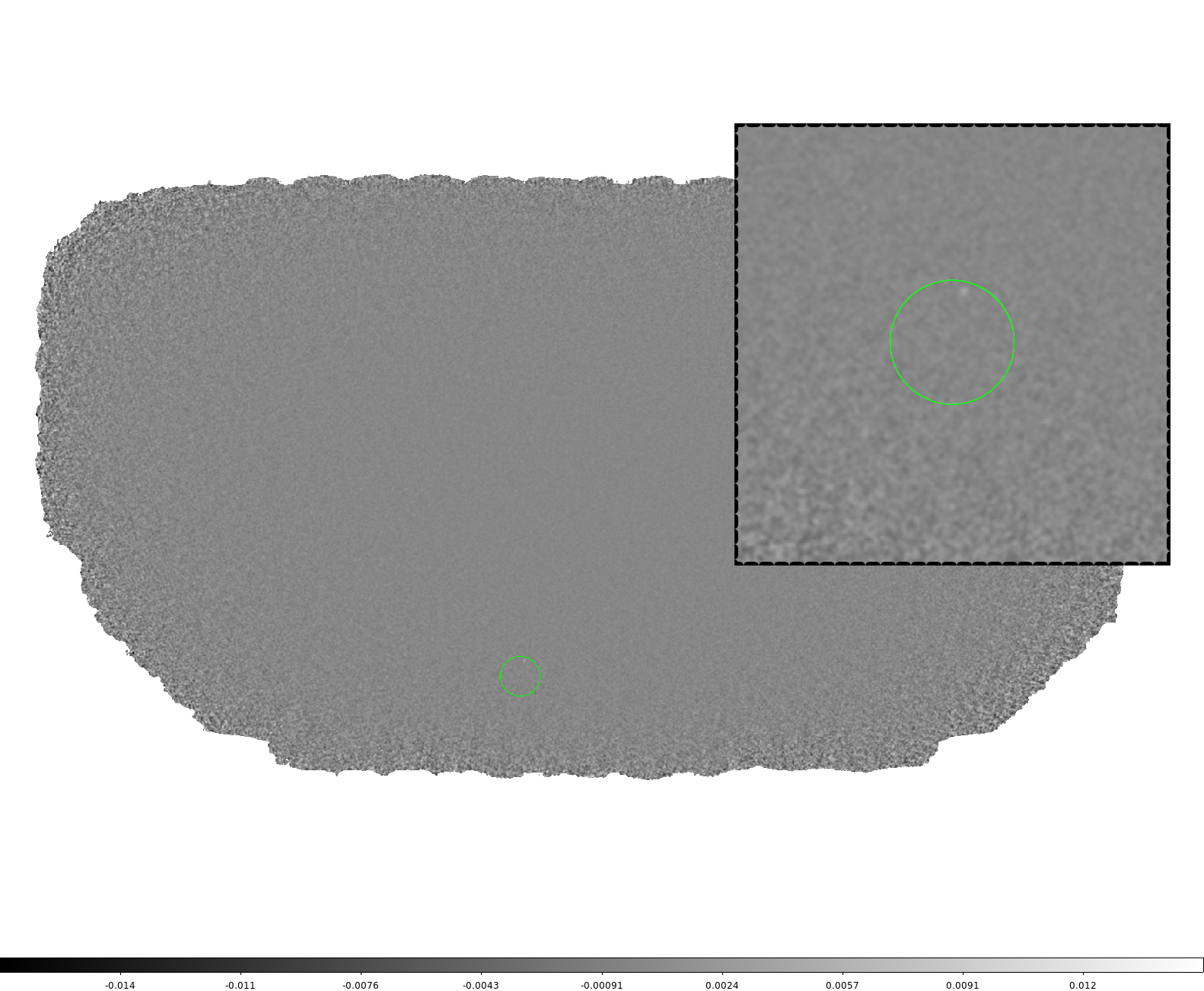}
	\caption{
	The sky-map from the Swift/BAT observation 00013484027 produced with the 
	batsurvey tool. 
	It shows the detected point source (white spot) that lies within 
	the error region of the BALROG position (light green circle) of the transient Vela X-1. 
	The color bar shows the intensity in units of counts/s.
  }
	\label{fig:velax1_swift_detection}
\end{figure}

To pinpoint the location and thereby establish the identification of this source we 
used the survey data of the {\it Swift}/BAT instrument.
During this activity the {\it Swift}/BAT survey has performed multiple observations 
covering this source region.
As {\it Swift}/BAT is more sensitive than GBM, the 5\,min integration time of BAT survey
mode should yield a detectable point source.
These observations were processed using the batsurvey tool \citep{HEAsoft2014},
which directly produces a sky-map.
Analysis of the observation 00013484027 lead to the detection of a point source at
the position of $\textrm{RA}=135.5342 {\rm \;deg}$ and $\textrm{Decl.}=-40.5619 {\rm \;deg}$ 
with a S/N of 28.1 (see Fig. \ref{fig:velax1_swift_detection}).
The catalog of known X-ray sources reveals this source to be Vela X-1.
Vela X-1, detected already with sounding rockets in the 1960ies \citep{Chodil+1967}, 
and contained in the Uhuru satellite catalog as 4U 0900-40 \citep{Forman+1978}, 
is a pulsing, eclipsing high-mass X-ray binary (HMXB) system that contains a neutron 
star and the supergiant star HD 77581 and is known for high energy flares \citep{Kretschmar2019}.
The neutron star has a spin period of 283.53\,s \citep{Kreykenbohm2008} which is 
nicely seen even in the raw data (Fig. \ref{fig:velax1_trigger_n6}) and consistent 
with our periodicity of 142\,s (we see emission from the north and the south pole of 
the neutron star separately).
Due to inhomogeneities in the stellar wind this accretion rate is not constant 
leading to the observed transient nature of its overall flux.
This remarkable result demonstrates the power of this novel transient detection
algorithm.

\section{Conclusions}

The main objective of this study was to develop a novel method for the detection 
of untriggered long-duration transient events in the data of {\it Fermi}-GBM, longer 
than the 5--10 min of the time scale of the background variations.
We applied a Bayesian framework to fit the physical background model, which
allows for an effective decoupling of transient sources from the complex and
varying background of the instrument.
The proposed trigger algorithm combines the signal of multiple detectors and
energy channels, providing an increased sensitivity in the detection of
change-points.
This new trigger algorithm is embedded into a data analysis pipeline that
continuously searches the data stream of the satellite and automates the
detection, localization, analysis and reporting of new transients.
This work establishes a procedure of automatic detection of slowly rising, 
long-duration transients in {\it Fermi}-GBM data, and pinpointing the location 
with an imaging instrument like {\it Swift}/BAT in survey mode.
The result of our extensive simulations gives evidence for a near-optimal 
(close to 100\% detection probability at $5\,\sigma$) trigger performance for long-duration signals.
This allowed for the establishment of the source flux limit as a function of transient duration, required for detection.
The fully automatic data analysis pipeline once enabled, allows the immediate 
use of this method for transient detections in the continuous data stream from the 
{\it Fermi}-GBM detector.

As a test example, we looked more closely at a triple trigger on June 17, 2020,
and could identify it with enhanced X-ray emission from the known Vela X-1 pulsar.

The improvements made to the background model in form of multi-detector fits, 
new sampling methods, and adjusted parametrizations has not only enabled the
transient detection in this work.
It could also significantly contribute to future applications of the physical background model 
such as the measurements of individual model parameters over long time-spans.
The sensitivity for new transient detections could be improved by combining the search 
with other observations as is proposed by the ``Astrophysical Multimessenger Observatory Network'' (AMON).
Joint detections of sub-threshold events in multiple instruments could significantly 
increase their plausibility.

The advances made in this study contribute to a greater understanding of long-duration
transients and nurture the analyses in the multi-messenger era.

\begin{acknowledgements} 
BB acknowledges support from the German Aerospace Center (Deutsches Zentrum f\"ur 
Luft- und Raumfahrt, DLR) under FKZ 50 0R 1913. 
JMB  acknowledges support from the Alexander-von-Humboldt Foundation. 
\end{acknowledgements}

\bibliographystyle{aa} 
\bibliography{library}

\end{document}